\documentclass{PoS}

\title{Results on the disconnected contributions for hadron structure}

\ShortTitle{Disconnected contributions}

\author{Constantia Alexandrou\\
        Department of Physics, University of Cyprus, P.O. Box 20537, 1678 Nicosia, Cyprus\\
        Computation-based Science and Technology Research Center, Cyprus Institute, 20 Kavafi Str., Nicosia 2121, Cyprus\\
        E-mail: \email{alexand@ucy.ac.cy}}

\author{Martha Constantinou\\
        Department of Physics, University of Cyprus, P.O. Box 20537, 1678 Nicosia, Cyprus\\
        E-mail: \email{marthac@ucy.ac.cy}}

\author{Vincent Drach\\
        CP3 Origins \& DIAS, University of Southern Denmark, Campusvej 55, DK-5230 Odense M, Denmark\\
        E-mail: \email{drach@cp3.dias.sdu.dk}}

\author{Kyriakos Hadjiyiannakou\\
        Department of Physics, University of Cyprus, P.O. Box 20537, 1678 Nicosia, Cyprus\\
        E-mail: \email{ph07hk2@ucy.ac.cy}}

\author{Karl Jansen\\
        NIC, DESY, Platanenallee 6, D-15738 Zeuthen, Germany\\
        E-mail: \email{Karl.Jansen@desy.de}}

\author{Giannis Koutsou\\
        Computation-based Science and Technology Research Center, Cyprus Institute, 20 Kavafi Str., Nicosia 2121, Cyprus\\
        E-mail: \email{g.koutsou@cyi.ac.cy}}

\author{\speaker{Alejandro Vaquero}\\
        Computation-based Science and Technology Research Center, Cyprus Institute, 20 Kavafi Str., Nicosia 2121, Cyprus\\
        E-mail: \email{a.vaquero@cyi.ac.cy}\footnote{Present address: Alejandro.Vaquero@mib.infn.it, INFN Sezione Milano Bicocca.}}

\abstract{We present results on the disconnected contributions to three point functions entering in studies of hadron structure. We use
$N_f=2+1+1$ twisted mass fermions and give a detailed description on the results of the nucleon sigma-terms, isoscalar axial charge and
first moments of bare parton distributions for a range of pions masses. In addition we give the $\sigma$-terms and the computations are
performed using QUDA code implemented on GPUs.}

\FullConference{The 32nd International Symposium on Lattice Field Theory,\\
		23-28 June, 2014\\
		Columbia University New York, NY}

\begin{document}

\section{Introduction}

Most studies of hadron structure have neglected the disconnected quark loop contributions due to the technical difficulty associated with
the computation of these  diagrams. Algorithmic developments as well as new computer architectures  are making the computation of
disconnected contributions feasible and enable us to reach the high precision necessary for obtaining meaningful results on these
quantities. In this work we employ GPUs for the complete evaluation of the disconnected  quark loops that contribute to the nucleon
observables~\cite{method}. Our implementation on GPUs includes the inversion of the Dirac operator, contractions and Fourier transform.
To this end we make extensive use of the QUDA library~\cite{QUDA}.

\section{Variance reduction techniques}
The basic quantity that enters in the computation of disconnected quark loops is the trace of the inverse of the fermionic matrix. Since
a direct computation is unaffordable, stochastic techniques are used (recently an alternative approach was developed based on
hierarchical probing~\cite{hProb,stefan}), to estimate the inverse matrix by employing stochastic noise sources~\footnote{In this work
we use $Z_N$ noise, as it was reported to be optimal~\cite{noise}.}. But stochastic techniques have inherent noise that decreases as
$O(1/\sqrt{N_r})$, with $N_r$ the number of stochastic sources employed. In addition, disconnected quark loops are prone to large gauge
noise and therefore variance reduction techniques are essential to obtain a good signal.

\subsection{The Truncated Solver Method}
In this work we  employ the well known Truncated Solver Method (TSM) ~\cite{TSM} that is found to be well-suited in
disconnected diagram computations due to its effectiveness and its low cost. The basics of the TSM consist of computing a cheap,
low-precision estimation of the inverse matrix by truncating the inverter using many stochastic noise sources, and then correcting it by a few
high-precision inversions. The correction is estimated by inverting a few noise sources to both high- and low-precision, and averaging over
the difference as follows

\begin{equation}
M_{E_{TSM}}^{-1}:= \frac{1}{N_{\rm HP}}\sum_{r=1}^{N_{\rm HP}}\left[\left|s_r\right\rangle_{\rm HP} - \left|s_r\right\rangle_{\rm LP}\right]\left\langle\eta_r\right| + \frac{1}{N_{\rm LP}}\sum_{j=N_{\rm HP}}^{N_{\rm HP}+N_{\rm LP}}\left|s_r\right\rangle_{\rm LP}\left\langle\eta_r\right|x.
\label{estiTSM}
\end{equation}
If the high- and the low-precision inversions yield results that are highly correlated, the correction does not fluctuate excessively,
and a few sources are enough to correct the bias. One then can use computer resources to increase the number of the low-precision
estimators, reducing the statistical error cheaply.

\subsection{The one-end trick}
The twisted-mass fermion regularization scheme (tmQCD) has a unique feature that allows for the use of a very powerful variance reduction technique for the
disconnected contributions~\cite{vvTrick}. By using the identities

\begin{eqnarray}
M^{-1}_u - M^{-1}_d = -2i\mu aM_d^{-1}\gamma_5M_u^{-1},\\
M^{-1}_u + M^{-1}_d = 2D_W,
\label{vvTrick}
\end{eqnarray}
any computation featuring a sum/difference of propagators is automatically improved. Note that we only need to replace the fields in
the contraction by

\begin{eqnarray}
\frac{2i\mu a}{N_r}\sum_{r=1}^{N_r} \left\langle s^\dagger_r \gamma_5 X s_r\right\rangle = \textrm{Tr}\left(M_u^{-1}X\right) - \textrm{Tr}\left(M_d^{-1}X\right) + O\left(\frac{1}{\sqrt{N_r}}\right),\\
\frac{2}{N_r}\sum_{r=1}^{N_r} \left\langle s^\dagger_r \gamma_5 X\gamma_5 D_W s_r\right\rangle = \textrm{Tr}\left(M_u^{-1}X\right) + \textrm{Tr}\left(M_d^{-1}X\right) + O\left(\frac{1}{\sqrt{N_r}}\right).
\label{loopVv}
\end{eqnarray}
This trick reduces the signal-to-noise ratio of the computation from $O\left(\frac{1}{\sqrt{V}}\right)$ to $O\left(1\right)$.



\section{GPU acceleration}

The QUDA library is used throughout for these  calculations with the exception of the generation of the sources, that is carried out on
CPUs. Once the source is copied to the GPU, the rest of the computations are performed there. Taking advantage of the specific
capabilities of the GPUs, the high-precision correction was computed using a mixed double-single precision solver, performing at 
$\approx 100$ GFlops per GPU, whereas the low-precision estimation used a mixed double-half precision solver, yielding the impressive
amount of $\approx 300$ GFlops per card. The contraction kernels in double precision achieve a performance of around 300 GFlops, solving
one of the most important bottlenecks that slowed down the loop computations on GPUs for some time. These kernels, developed by our
group~\cite{yoGPU}, deliver results for all possible $\gamma$ insertions for ultra-local and one-derivative operators, which allowed us
to perform a complete analysis of all disconnected loop contributions to hadron structure. The Fourier transform required was
performed by employing the highly optimized CUDA library cuFFT. These improvements enable  a very efficient computation of the
disconnected quark loops.

\section{Ensembles}

We show results for two different tmQCD $N_F=2+1+1$ ensembles:

\begin{table}[h!]
\begin{center}

\begin{tabular}{c|ccccc}
Ensemble &		Volume		& $m_\pi$ (MeV) & $m_\pi L$ &  $a$ (fm)  & Stats  \\
\hline 
B55.32   & $32^3\times 64$	&	372(5)		&	4.97	& 0.0823(10) & 147072 \\
D15.48   & $48^3\times 96$	&	213(21)		&	3.35	& 0.0646(7)  & 7752
\end{tabular}

\caption{Ensembles used in the calculation.}\label{results}
\end{center}
\end{table}

\noindent
Both the strange and the charm quark masses were tuned to their physical values for both ensembles. The low-precision inversions used
a residual of  $\rho\approx 5\times 10^{-3}$, while the high-precision inversion used  $\rho\approx 10^{-9}$. For  the D15.48 ensemble
we computed 24 high-precision propagators for the correction and 500 low-precision ones for the estimation for all flavors. For the B55.32 we also use 24
high-precision propagators and 500 low-precision ones for the light quark sector, while for the strange and the charm quark sectors  300 low-precision propagators are used.

\section{Results on nucleon observables}

\begin{figure*}[h!]
\begin{center}
\begin{minipage}{0.45\linewidth}
\includegraphics[width=\linewidth,angle=0]{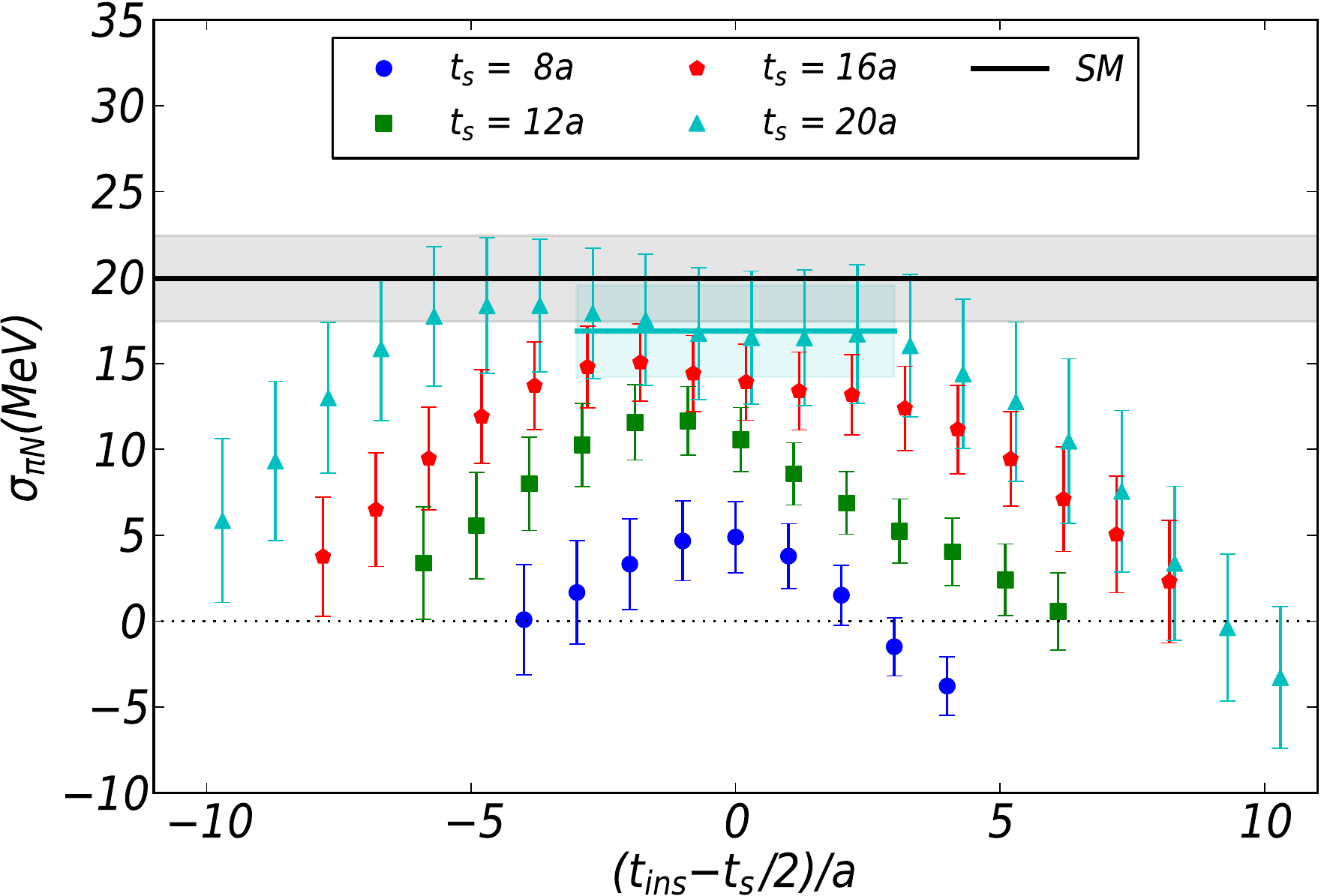}\\
\includegraphics[width=\linewidth,angle=0]{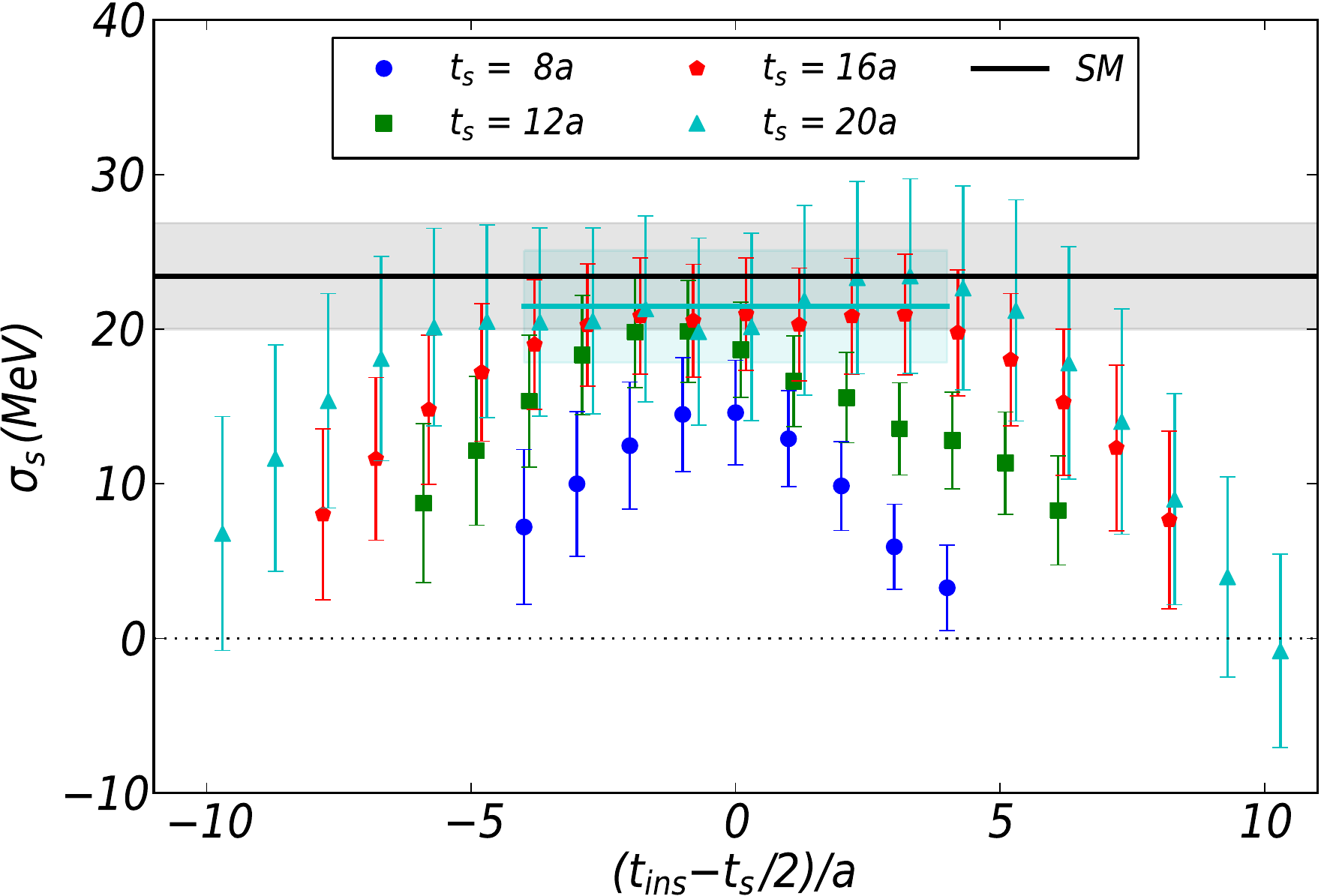}
\end{minipage}
\begin{minipage}{0.45\linewidth}
\includegraphics[width=\linewidth,angle=0]{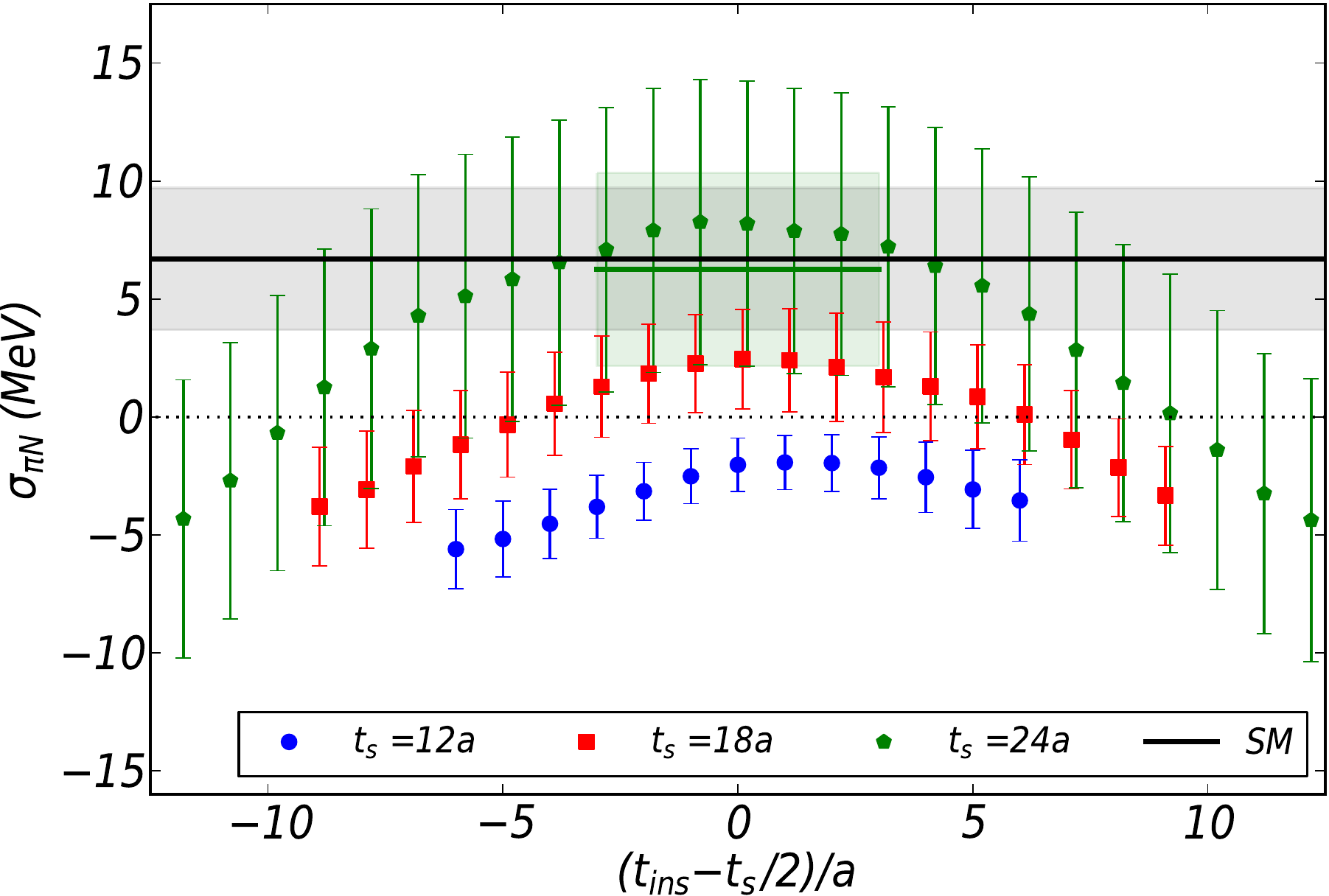}\\
\includegraphics[width=\linewidth,angle=0]{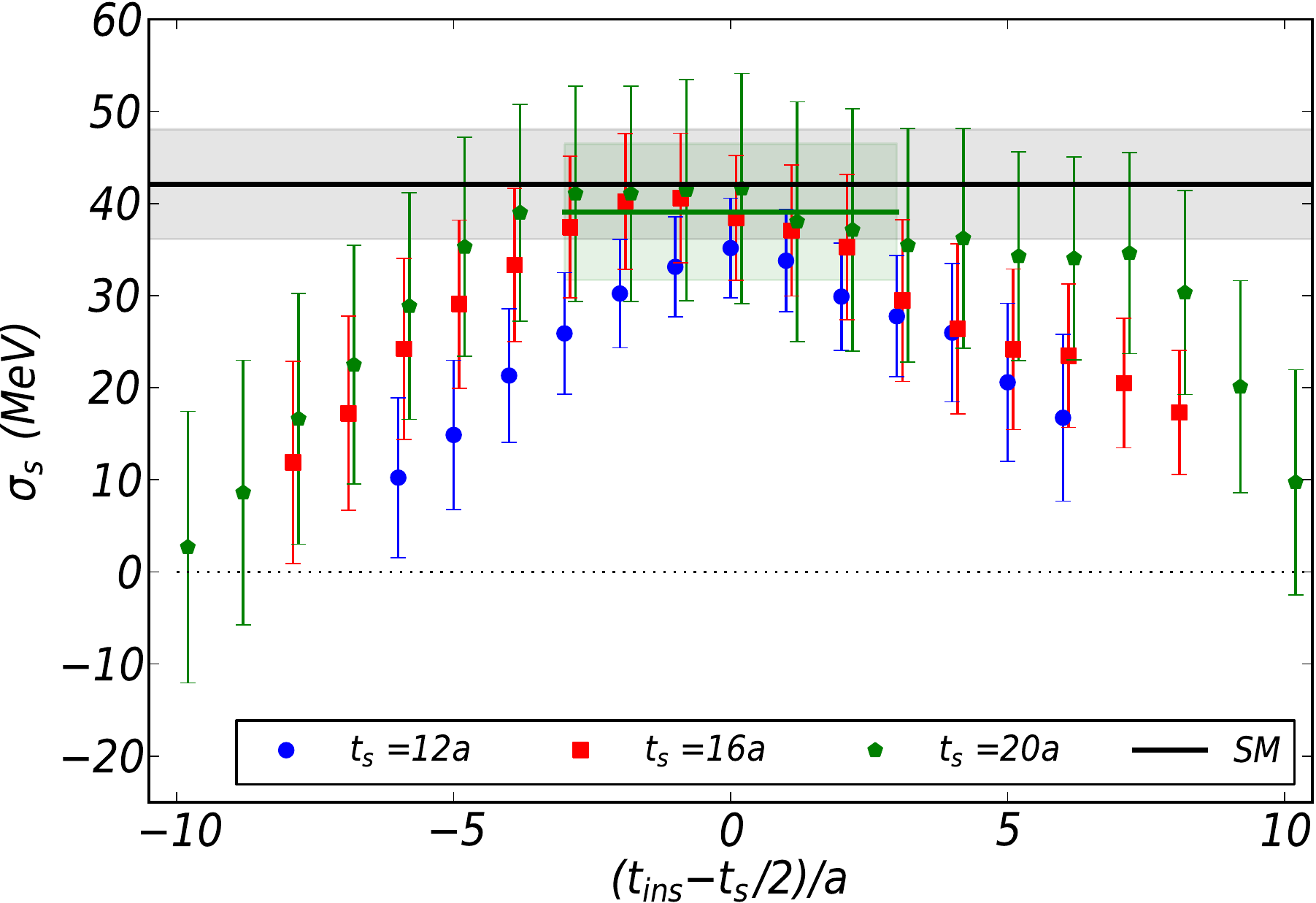}
\end{minipage}
\caption{Plateau method for the nucleon $\sigma_{\pi N}$ (upper panel) and  $\sigma_s$ (lower panel) for the B55.32 (left) and D15.48
(right) ensembles. Contamination from excited states is evident, as the value of the plateau increases with increasing $t_s$ up
to $t_s\sim 1.5$~fm. The grey band is the result obtained from the summation method.\label{Sigma}}
\end{center}
\vspace{-0.4cm}
\end{figure*}

We show results on the bare nucleon $\sigma$--terms in Fig.~\ref{Sigma} that describe the scalar content of the nucleon, and are highly
relevant in dark matter searches. Comparing the results of both ensembles we appreciate a clear decrease in the importance of the
contribution of $\sigma_{\pi N}$ disconnected, whereas $\sigma_s$ grows with diminishing mass. The behavior of $\sigma_c$ with the pion
mass is not clear to us, for from the data coming from the B55 ensemble we couldn't find a reasonable value.

Using these two ensembles and a linear extrapolation in $m_\pi^2$ we find (for the disconnected, no connected contributions here) 
$\sigma_{\pi N} = 3.9\pm 4.5$ MeV and $\sigma_s = 47.8 \pm 8.7$ MeV at the physical point. Given that we only have two ensembles, with
limited statistics for the D15.48, the values of the $\sigma$-terms quoted are to be regarded as preliminary. The important point is
that we have a method that with increased statistics can be applied to the computation of these important observables.

\begin{figure*}[h!]
\begin{center}
\begin{minipage}{0.45\linewidth}
\includegraphics[width=\linewidth,angle=0]{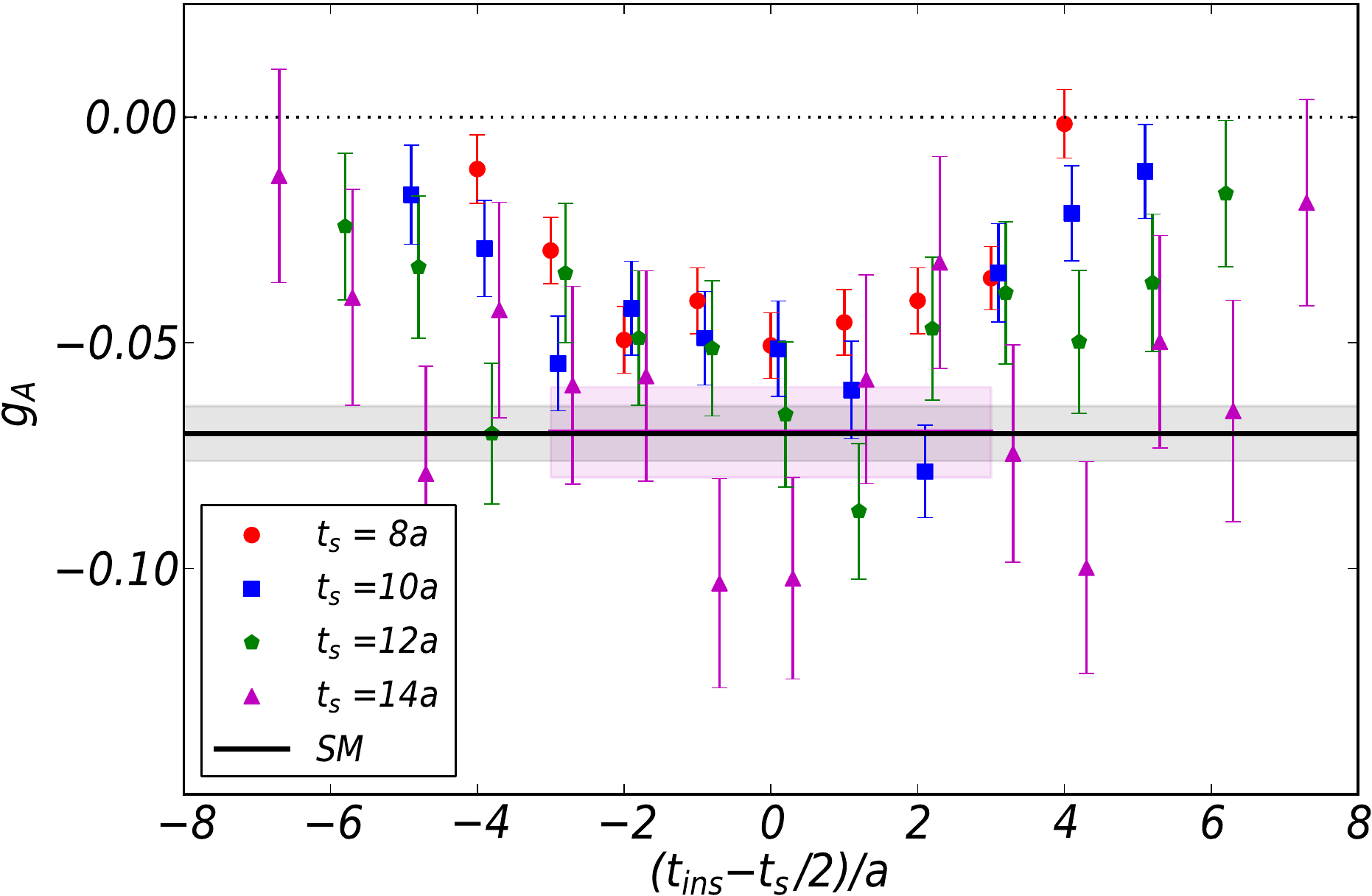}\\
\includegraphics[width=\linewidth,angle=0]{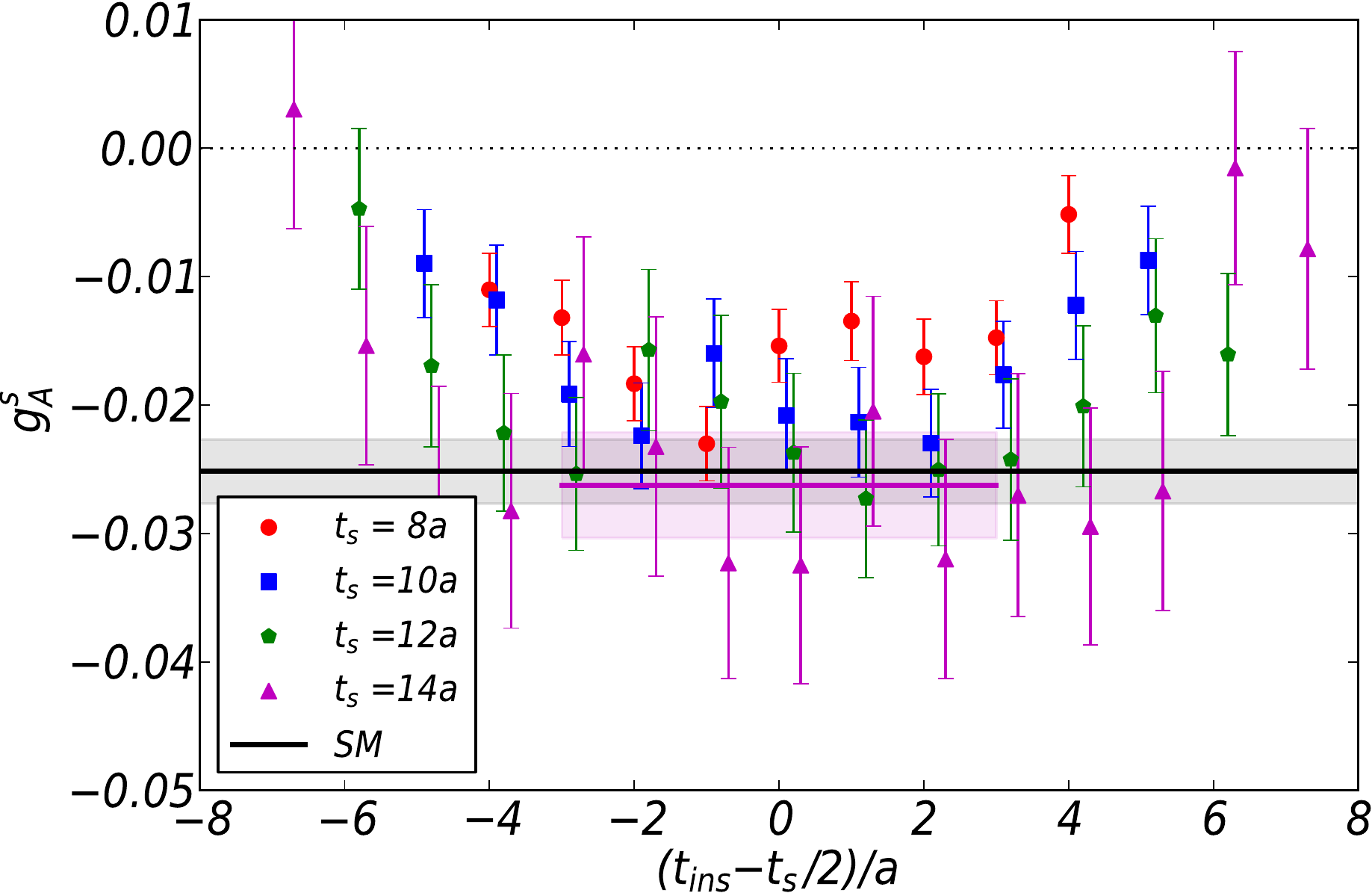}
\end{minipage}
\begin{minipage}{0.45\linewidth}
\includegraphics[width=\linewidth,angle=0]{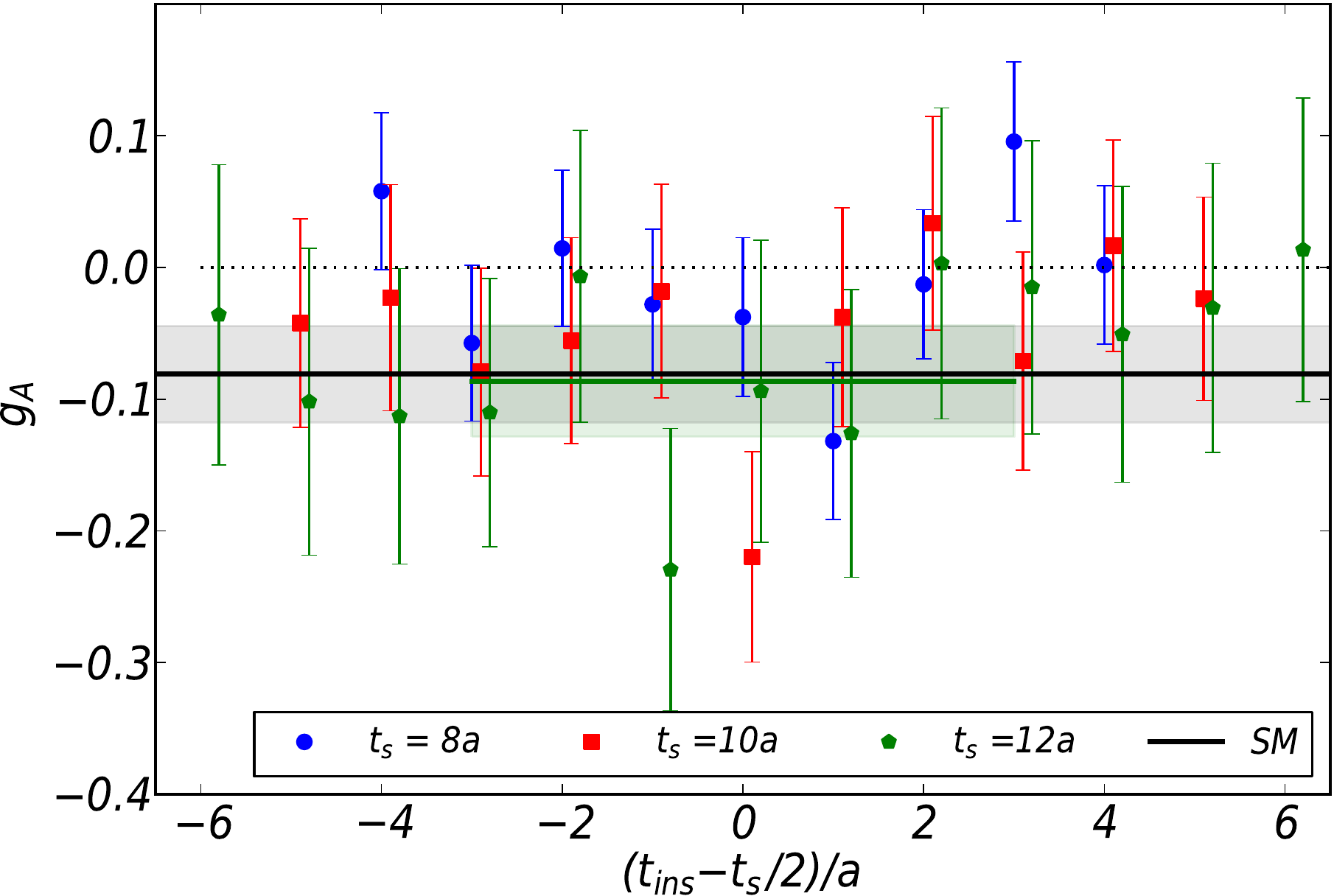}\\
\includegraphics[width=\linewidth,angle=0]{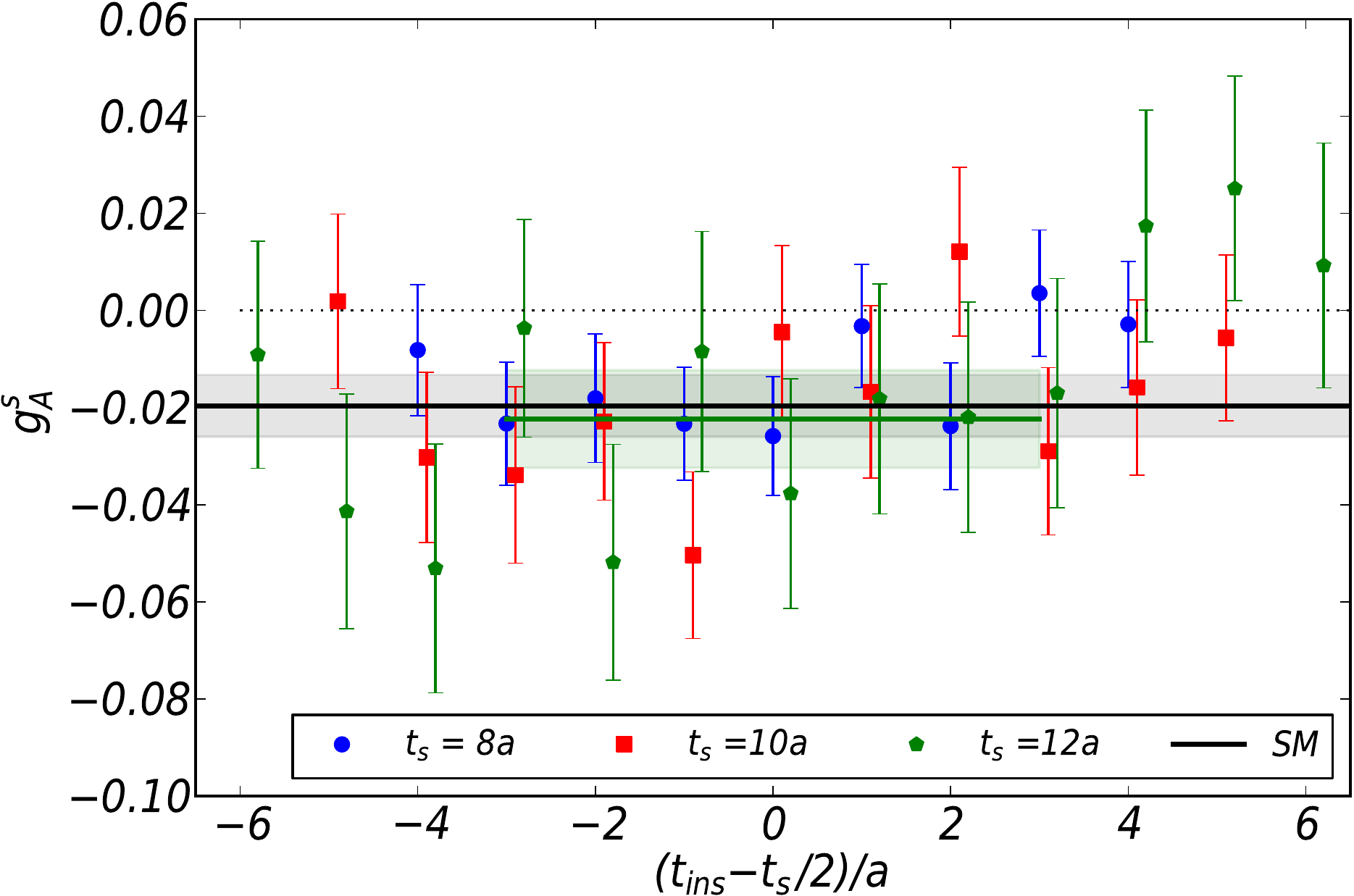}
\end{minipage}
\caption{Plateau method for the bare nucleon $g_A^{u+d}$ (upper panel) and $g_A^s$ (lower panel) for the B55.32 (left) and D15.48 (right)
ensembles. No noticeable contamination from excited states is observed.\label{gAL}}
\end{center}
\vspace{-0.4cm}
\end{figure*}

The value of the axial charge $g_A^q$ determines the intrinsic fraction of the spin carried by a quark $q$ in the proton. Given
the long-standing spin puzzled of the nucleon it is important to be able to compute this quantity directly from QCD. To extract $g_A^q$
we need both the isovector and isoscalar values of the matrix element of the axial-vector current. In Fig.~\ref{gAL} we show the
results on the disconnected part of the isoscalar\footnote{For the renormalization we use the isovector $Z_A$. The error introduced is
of order $O(g_0^2)$, much smaller than our current statistical errors.} $g_A^{u+d}$ and strange contribution to the spin $g_A^s$.
\noindent
If one extrapolates to the physical pion mass to LO (constant fit), one obtains $g_A^{u+d} = -0.075\pm 0.038$ and
$g_A^s = -0.0212\pm 0.0072$. While these numbers are preliminary it is clear that the correction to the connected part is
$\approx 10\%$, and therefore disconnected contributions must be taken into account in determining high precision results for the spin
content of the nucleon.

\begin{figure*}[h!]
\begin{center}
\begin{minipage}{0.45\linewidth}
\includegraphics[width=\linewidth,angle=0]{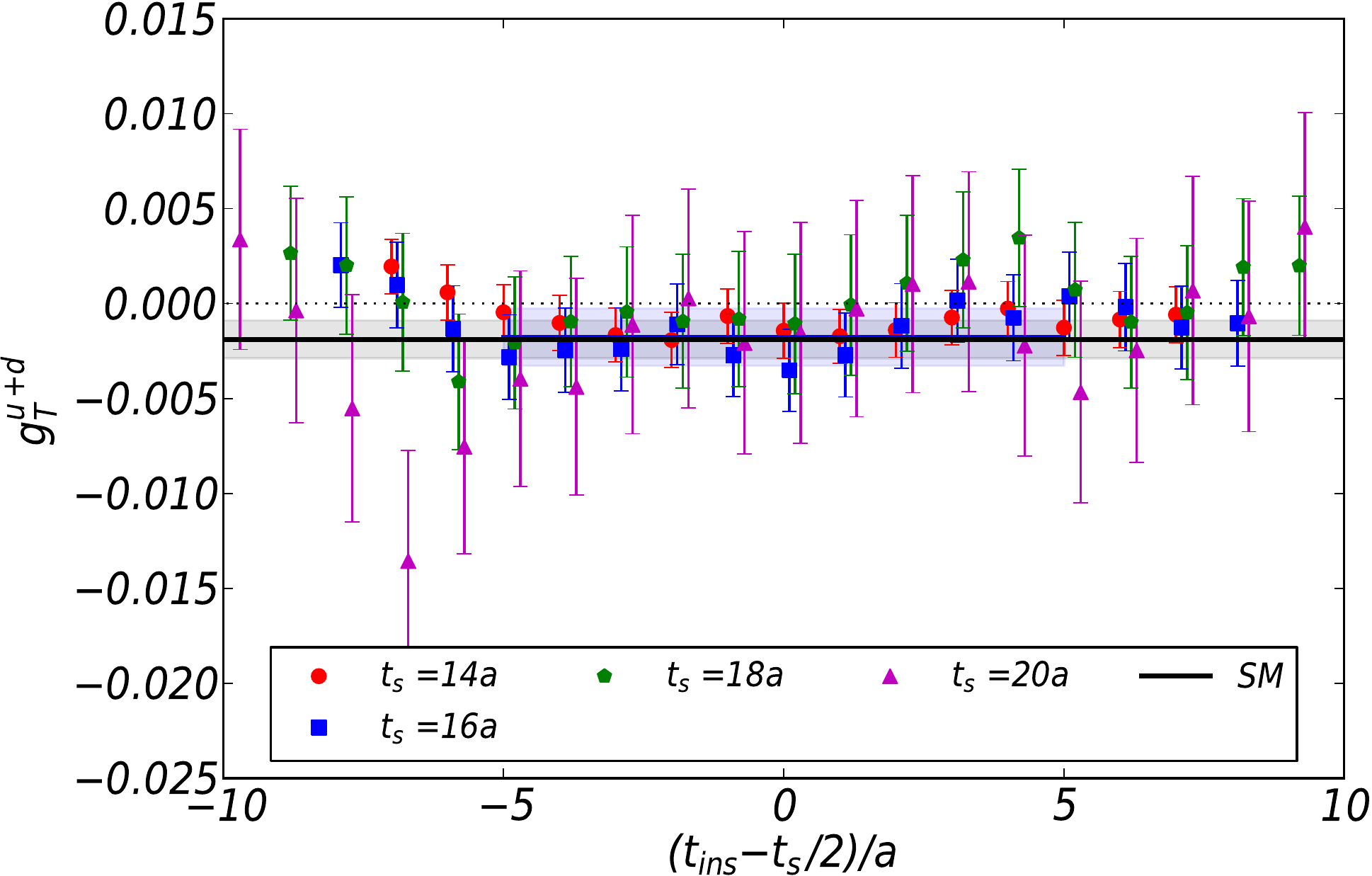}\\
\includegraphics[width=\linewidth,angle=0]{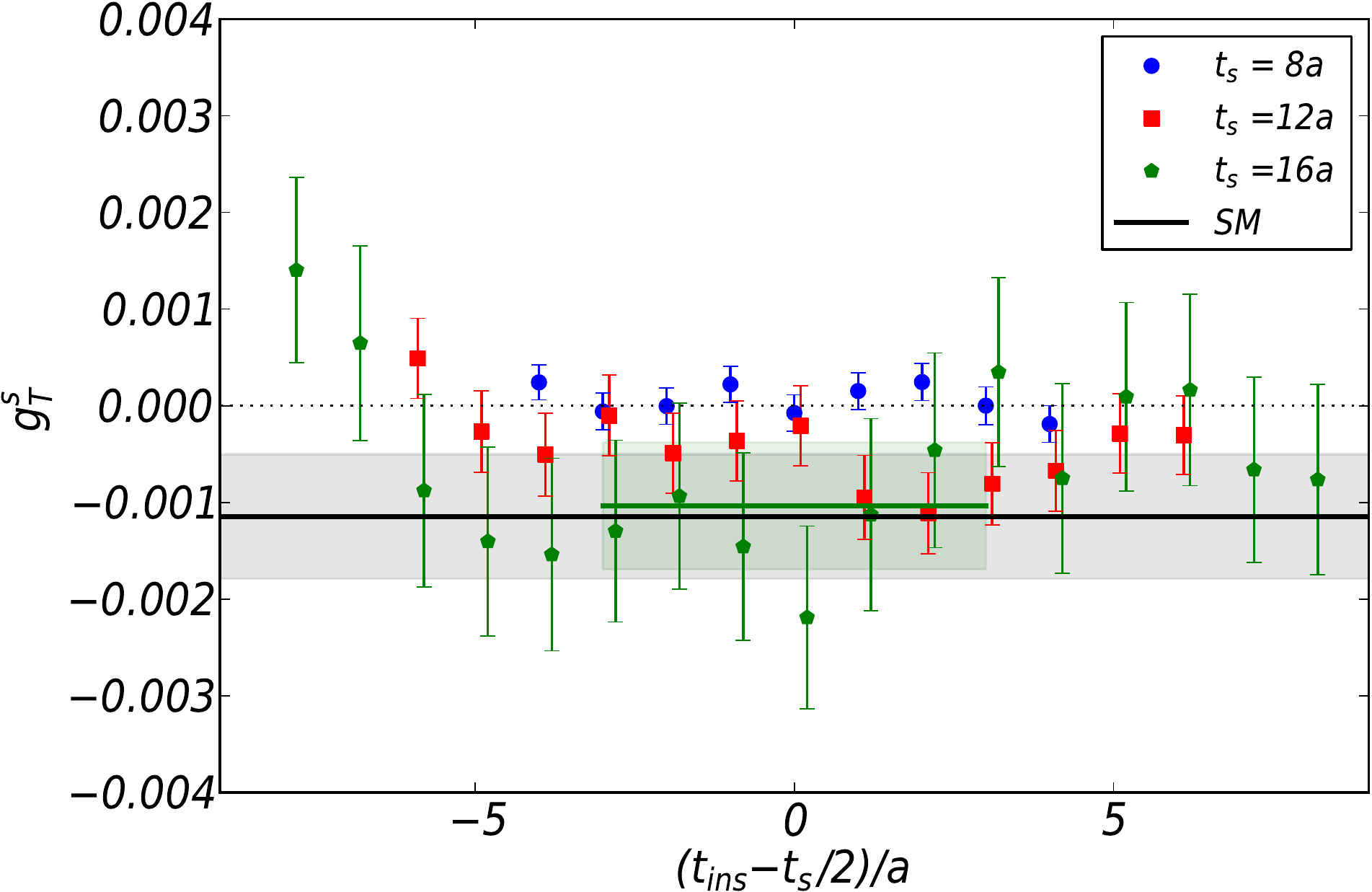}
\end{minipage}
\begin{minipage}{0.45\linewidth}
\includegraphics[width=\linewidth,angle=0]{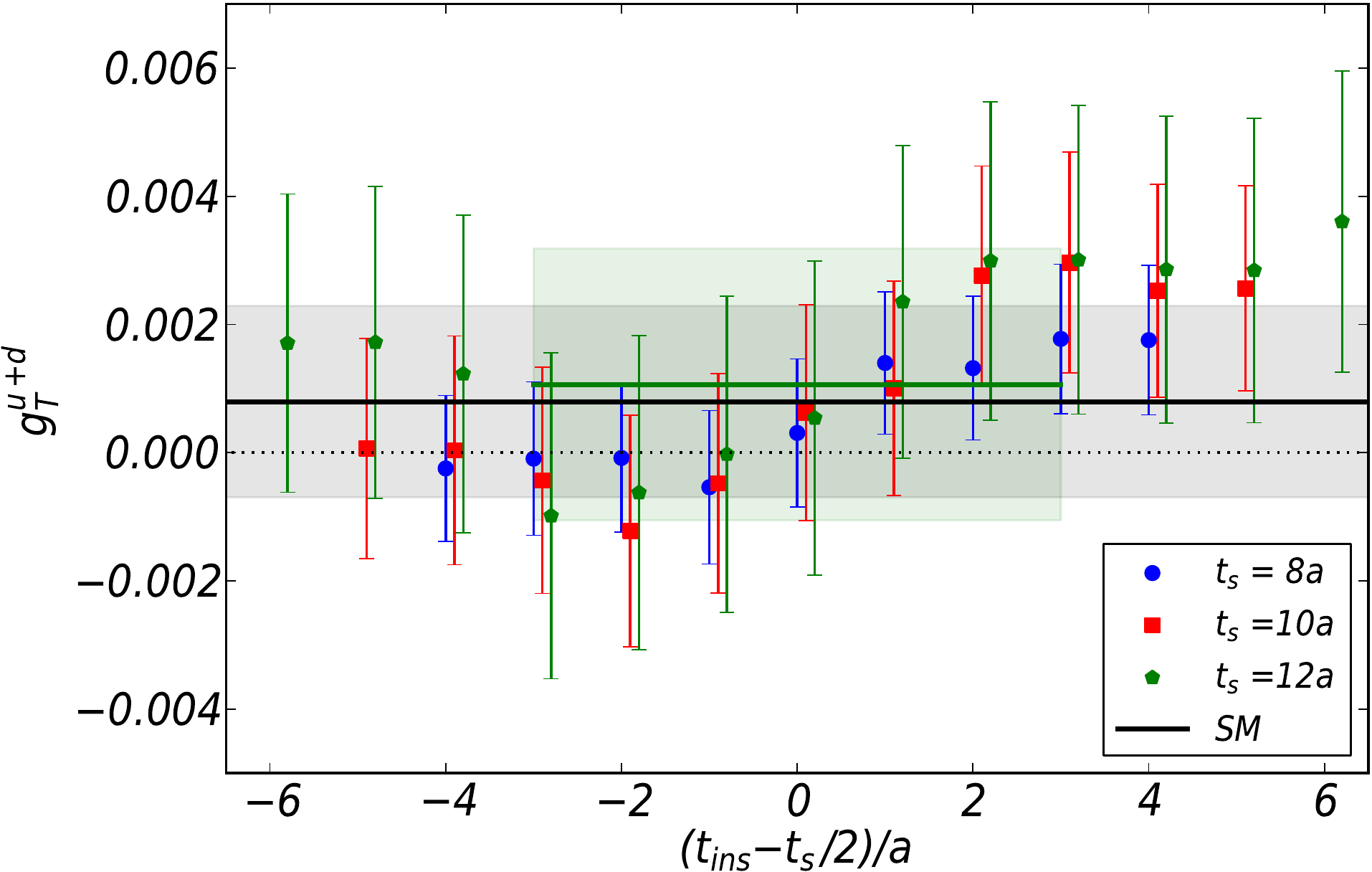}\\
\includegraphics[width=\linewidth,angle=0]{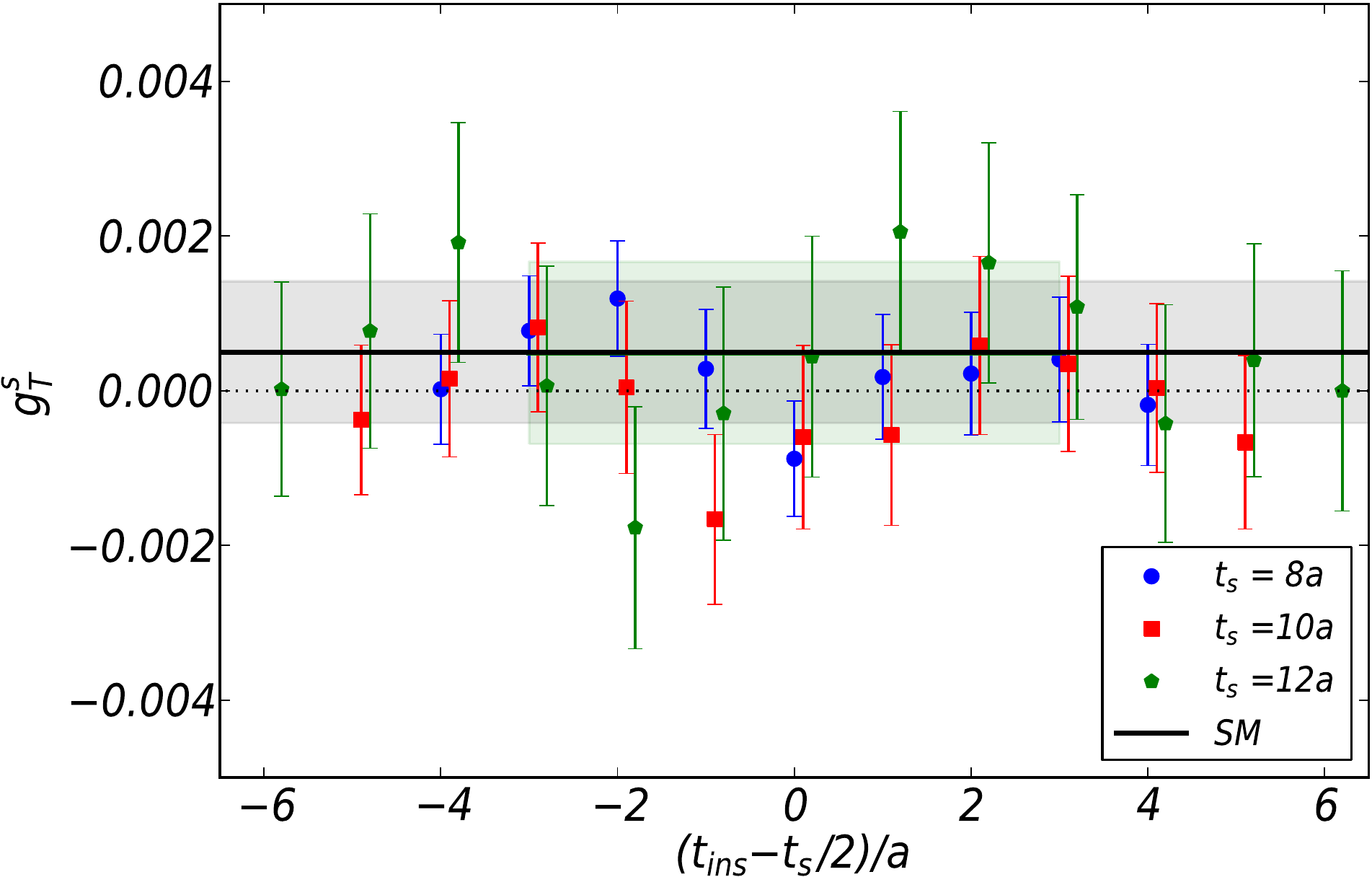}
\end{minipage}
\caption{Plateaux for $g_T^{u+d}$ (upper panel) and $g_T^s$ (lower panel) of the nucleon for the B55.32 (left) and the D15.48 (right) ensembles.\label{gTplot}}
\end{center}
\vspace{-0.4cm}
\end{figure*}

Our study includes loops with all $\gamma$-structure and thus also the tensor combination entering in the tensor content of the nucleon.
Contrary to $g_A^{u+d}$ our  results on $g_T^{u+d}$ displayed in Fig.~\ref{gTplot} show that the disconnected contribution is
very small and can be neglected compared to the connected contribution. This is also true for $g_T^s$.

\begin{figure*}[h!]
\vspace{-0.2cm}
\begin{center}
\begin{minipage}{0.3125\linewidth}
\includegraphics[width=\linewidth,angle=0]{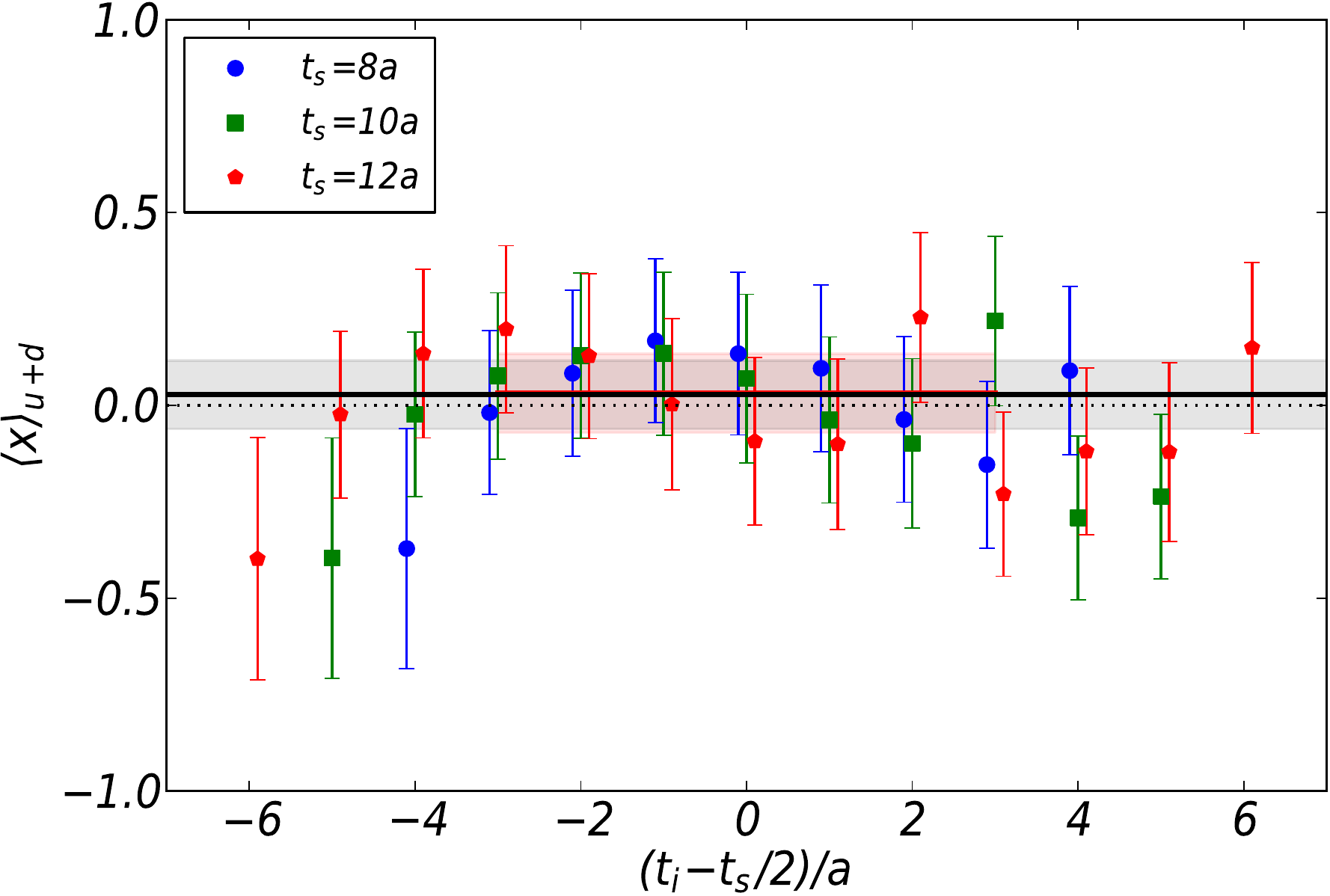}
\end{minipage}
\begin{minipage}{0.3125\linewidth}
\includegraphics[width=\linewidth,angle=0]{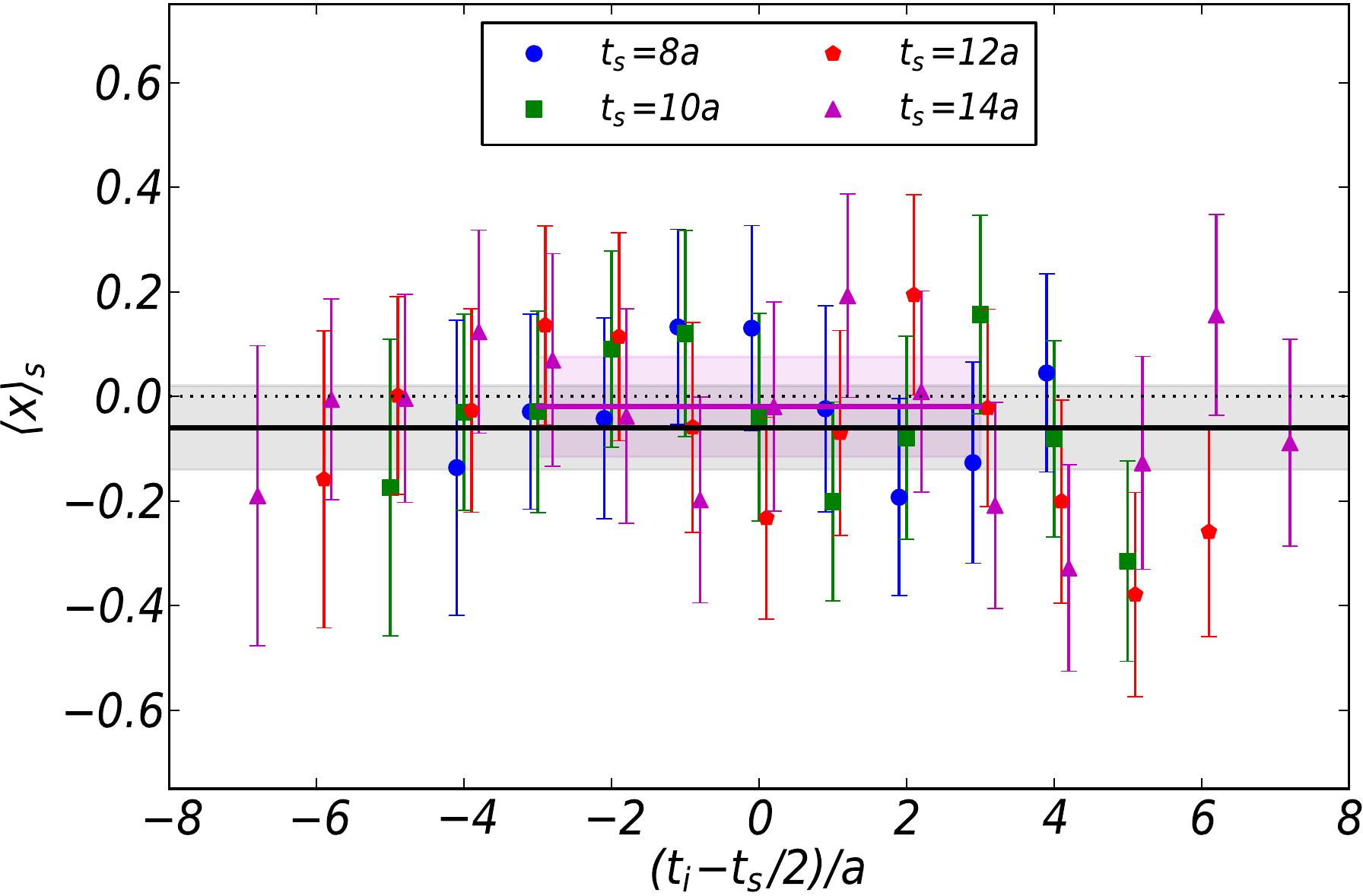}
\end{minipage}
\begin{minipage}{0.3125\linewidth}
\includegraphics[width=\linewidth,angle=0]{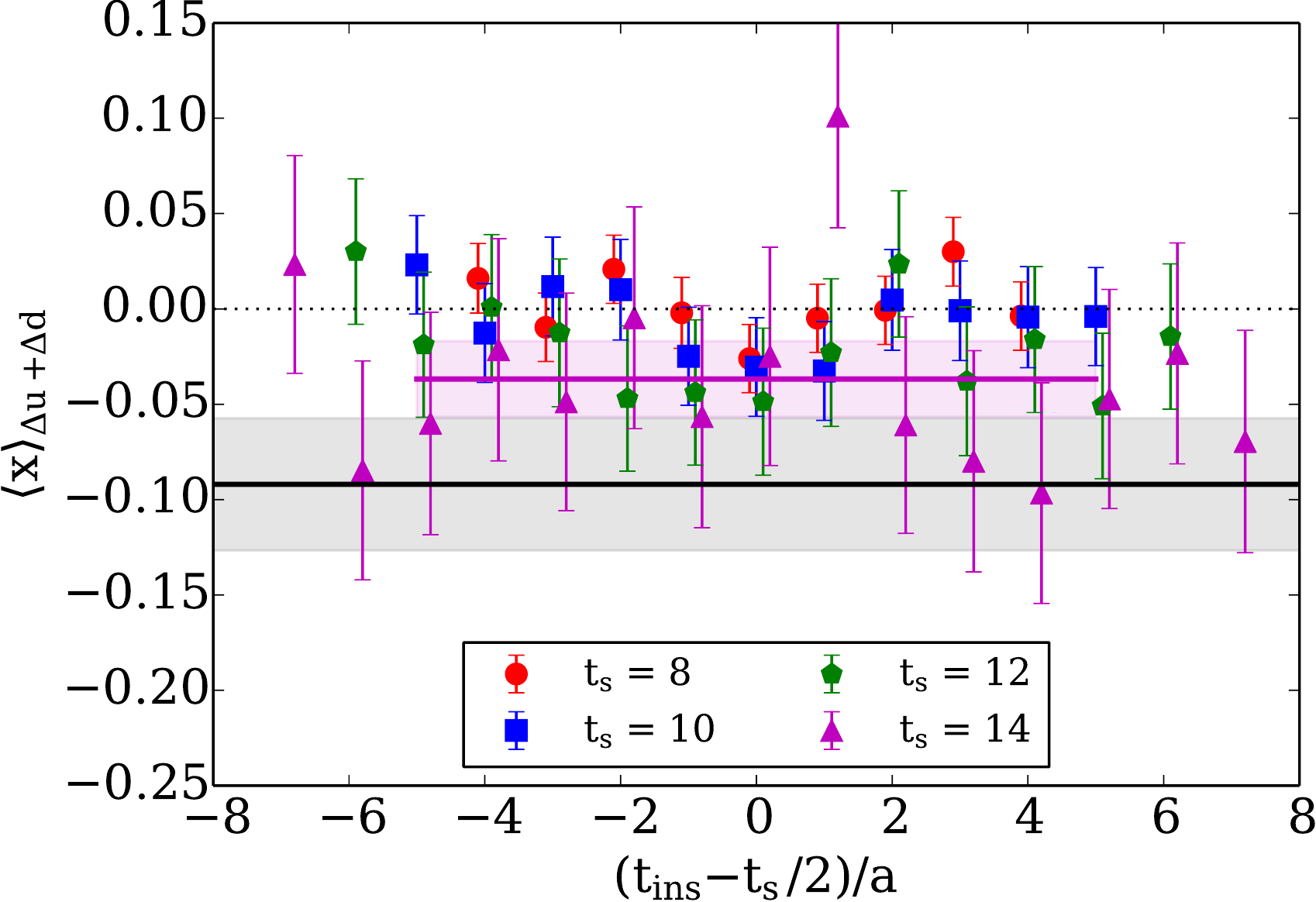}
\end{minipage}
\caption{Results for $\left\langle x\right\rangle_{u+d}$ (left), $\left\langle x\right\rangle_s$ (center) and 
$\left\langle x\right\rangle_{\Delta u + \Delta d}$ (right) of the nucleon for the B55.32 ensemble.\label{momHelB55}}
\end{center}
\vspace{-0.4cm}
\end{figure*}

Our results include disconnected contributions entering the momentum fraction carried by the light and the strange quarks in the nucleon.
Within our accuracy these are found consistent with zero for the B55.32 ensemble as shown in Fig.~\ref{momHelB55} for the momentum
fraction and helicity. Similar results are obtained for the D15.48 ensemble. 

\section{The physical point}

In order to eliminate the systematic error due to the chiral extrapolation one needs to compute the disconnected contributions
directly at the physical point. Although we have an ensemble of twisted mass fermion with a clover term at the physical point, applying
the methodology developed for larger pion masses proved to be insufficient. The reason is that the size of the correction involved in
the TSM algorithm grows as the mass decreases, requiring a large number of high-precision inversions (and hence becoming prohibitively
expensive) to obtain a reliable value for the quark loops. Reducing the residual of the low-precision estimation can help, but increases
the cost of the computation up to a level that the TSM is not efficient any more. Attempts to tune for the best values following the
procedure outlined in~\cite{TSM} failed to give parameters that would make the TSM competitive so far. Since the low-modes of the Dirac
operator are expected to be responsible for this behavior, we are exploring deflation techniques in order to remove them. There already
exists reports on the performance of the deflated TSM, usually called Truncated Eigenmode Acceleration (TEA)~\cite{TEA}.

\section{Conclusions}

The disconnected contributions are becoming finally accesible by a clever combination of computer power and state-of-the-art algorithms.
This allowed us to carry out a broad, high-precision study that included all the possible disconnected contributions with ultra-local
and one-derivative insertions. The final objective is to remove the systematic errors that appear in many hadron structure studies, as
well as calculate all these disconnected contributions that have interest by themselves (e.g. $\sigma_s$ for dark matter searches) to a
new level of precision.

In spite of the large improvements recently made in algorithms, high statistics are still required to obtain good signal in many
observables. The most accurate results presented here were for $\approx 150000$ measurements, and even at such high statistics we were
not able to achieve the accuracy typically achieved with connected contributions for all observables.

In the future we plan to deliver data obtained directly at the physical pion mass. To this end we will include deflation in our code
and calculate the low-modes exactly.

\section{Acknowledgements}
A. Vaquero and M. Constantinou  are supported by the Cyprus Research Promotion Foundation (RPF) under contracts EPYAN/0506/08 and
TECHNOLOGY/$\Theta$E$\Pi$I$\Sigma$/0311(BE)/16 respectively and K. Jansen partly by the RPF project
$\Pi$PO$\Sigma$E$\Lambda$KY$\Sigma$H/EM$\Pi$EIPO$\Sigma$/0311/16. This work was partly funded by the RPF project 
NEAY$\Pi$O$\Delta$OMH/$\Sigma$TPATH/0308/31 (infrastructure project Cy-Tera, co-funded by the European Regional Development Fund and
the Republic of Cyprus through the RPF). Computer resources were provided by the Cy-Tera machine of CaSToRC,
Forge at NCSA Illinois (USA), Minotauro at BSC (Spain), and Juqueen at the J\"ulich Supercomputing Center.

\end{document}